\documentclass[10pt, letterpaper,conference]{IEEEtran}

\usepackage{amsmath,amssymb,amsthm,epsfig,epstopdf}
\usepackage{float}
\usepackage[mathscr]{euscript}
\usepackage{multicol}
\usepackage[bottom]{footmisc}
\usepackage{subfigure}
\usepackage{multicol}
\usepackage{comment}

\usepackage{rotating}
\usepackage{epsfig}
\usepackage{epstopdf}
\usepackage[nospace,noadjust]{cite}
\usepackage{graphicx,dblfloatfix}
\usepackage{calrsfs}
\DeclareMathAlphabet{\pazocal}{OMS}{zplm}{m}{n}
\newcommand{\Lb}{\pazocal{N}}

\usepackage{color}

\usepackage[linesnumbered,ruled,vlined]{algorithm2e}
\usepackage{algpseudocode}

\algnewcommand{\algorithmicor}{\textbf{ or }}
\algnewcommand{\OR}{\algorithmicor}

\graphicspath{{Figures/}}
\newtheorem{lem}{Lemma}
\newtheorem{defn}{Definition}

\begin{document}

\title{Energy-Efficient Coalition Formation in Sensor Networks: a Game-Theoretic Approach}

\author{M. Mehdi Afsar$^1$, R. Trafford Crump$^2$, and Behrouz H. Far$^1$ 
\\ $^1$Department of Electrical and Computer Engineering,\\$^2$Cumming School of Medicine,\\University of Calgary\\{ (E-mail: afsar@ieee.org, \{tcrump, far\}@ucalgary.ca)} 
}

\maketitle

\begin{abstract}
The most important challenge in Wireless Sensor Networks (WSNs) is the energy constraint.  Numerous solutions have been proposed to alleviate the issue, including clustering. Game theory is an effective decision-making tool that has been shown to be  effective in solving complex problems. In this paper, we employ cooperative games and propose a new clustering scheme called Coalitional Game-Theoretic Clustering (CGTC) algorithm for WSNs.  The idea is to partition the entire network area into two regions, namely \textit{far} and \textit{vicinity}, in order to address the hotspot problem in WSNs, wherein nodes close to the base station (BS) tend to deplete their energy faster due to relaying the traffic load received from farther nodes. Then, coalitional games are utilized to group nodes as coalitions.  The main factor in choosing coalition heads is the energy level of nodes so that the most powerful nodes play the role of heads. The Shapley value is adopted as the solution concept to our coalitional games. The results of simulations confirm the effectiveness of CGTC in terms of energy efficiency and improved throughput.
\end{abstract}


\IEEEpeerreviewmaketitle

\section{Introduction}
\label{sec:intro}
Wireless sensor networks (WSNs) consist of a large number of tiny, smart sensor nodes.  WSNs are a building block in the Internet of Things (IoT) and their applications are widespread, ranging from military to healthcare to smart transport systems. Since the sensor nodes are usually battery-operated, the most important challenge is to conserve their energy.  Accordingly, researchers have proposed  a variety of solutions, trying to address this challenge. A popular solution is to cluster the sensor nodes.  With this solution, the nodes are divided into some groups known as clusters, with some nodes elected to play the role of cluster heads.  
 Clustering has many benefits for WSNs, including energy-efficiency, scalability, and topology management~\cite{afsar-survey}. 

 Game Theory (GT) is a framework to study interactions among intelligent rational players~\cite{Han:2012:GTW}.  Introduced by Neumann and Morgenstern~\cite{neumann1944}, the modern GT is utilized today in various fields, including  economics, politics, biology, and computer science.  
 In GT, each player has a set of choices, and it is assumed that each player plays rationally and selects the strategy  that maximizes their {\it utility} (or {\it payoff}).  In general, games could be divided into two groups: cooperative and non-cooperative. In cooperative games, players prefer network-wide utilities over individual ones. In non-cooperative games, each player plays individually and opts for decisions to maximize their own utility.   A previous work has reported that cooperative games have a better performance in group formation for WSNs than non-cooperative games~\cite{GT-survey2015}.
 
 The \textit{hotspot} problem appears in clustered WSNs, where nodes close to the base station (BS) relay the highest amount of load received from farther nodes.  The \textit{unequal} clustering, first proposed in~\cite{soro2005}, tries to form smaller clusters around the BS to cope with the unbalanced energy consumption problem.  

In this paper, we target the energy efficiency problem in WSNs by proposing Coalitional Game-Theoretic Clustering (CGTC) algorithm. First, the network area is divided into two regions: \textit{far} and \textit{vicinity}.  In the \textit{far} region, a set of nodes with the highest residual energy, called Coalition Head Nominees (CHNs), initiate cooperative games within their surroundings.  Then, CHNs along with some other nodes shape final coalitions.  CHNs select these nodes based on their distance to other neighbouring nodes and to the BS, explained in Section~\ref{sec:alg-det}. In the \textit{vicinity} region, some small coalitions are formed to tackle the energy-consuming data relaying task.  The Shapley value~\cite{shapley1952value} is chosen as the solution concept to our coalitional games.  

The rest of this paper is outlined as follows. Section~\ref{sec:rel-wo} provides a state-of-the-art survey on game-theoretic clustering algorithms for WSNs. Section~\ref{net-mod} explains the modelling and definitions, required for Section~\ref{sec:alg-det} which describes the proposed CGTC algorithm in detail. Analyzing the parameters of CGTC is performed in section~\ref{sec:anal}.  We present the simulation experiments in section~\ref{sec:per-eval}, and the paper is concluded in section~\ref{sec:con}.

\section{Related Work}
\label{sec:rel-wo}
 While clustering has been a popular mechanism in large scale computer networks to make them scalable~\cite{kleinrock1977hierarchical}, the first clustering protocol for WSNs was LEACH~\cite{LEACH2000}. LEACH is a random cluster head selection algorithm in which the role of cluster head is rotated among all nodes in a predefined time duration. LEACH was designed for a small scale WSN where the communication between sensor nodes and the BS was direct; nonetheless, in modern large scale WSNs, direct communication between sensor nodes and the BS is not energy-efficient and multi-hop communication should be adopted.
 The main problem in clustered WSNs, where multi-hop is adopted to relay the data from farther nodes toward the BS, is the hotspot problem and cluster heads close to the BS deplete their energy faster due to excessive data relaying. This problem was first addressed by Soro and Heinzelman~\cite{soro2005} where smaller clusters are formed in the vicinity of the BS, so the cluster heads can conserve more energy for the relaying task. We utilize this strategy in our scheme to solve the hotspot problem. Studying non-game-theoretic clustering algorithms for WSNs is not the focus of this paper, so the reader is referred to~\cite{afsar-survey} for a good review of clustering algorithms for WSNs. In the following, we specifically focus on algorithms that utilize GT to cluster WSNs.

Recently, a few clustering algorithms based on GT have been proposed, and they could generally be classified into: non-cooperative and cooperative clustering algorithms. DEGRA~\cite{DEGRA} is a non-cooperative clustering algorithm in which a finite complete and perfect information game is employed and the payoff of each player consists of three factors: the residual energy of the nodes, the average energy consumption of neighboring nodes, and the node density.  In GTC~\cite{GTC2012}, GT is used in order to tune the cluster sizes.  
Initially, the network area is segmented into some squares. Then, the width of each square is determined via GT in order to equalize the load among all Coalition Heads (CHs), considering the fact that there is exactly one CH in each square.  
Bascially, since each node tries to individually achieve a better payoff, non-cooperative games fail to be the best match for group formation purposes in WSNs~\cite{GT-survey2015}. 

On the other hand, quite a few clustering algorithms based on cooperative games have been proposed in recent years.  In CGC~\cite{intech2010}, a cooperative clustering algorithm is introduced for WSNs with the objective of maximizing the network lifetime. The idea behind CGC is to consider a trade-off between individual and network-wide costs. 
Formed coalitions consider the number of cluster members, the redundant energy, and the transmission energy.  The Shapley value is chosen as the solution that assigns a single cost allocation to the cost sharing game.  Nonetheless, initial candidate selection in CGC is random, which significantly diminishes its reliability.  Moreover, CGC has the scalability problem since communication between the nodes and the BS is direct.  We compare the performance of our algorithm with that of CGC in section~\ref{sec:per-eval}.  Similarly, CSGC~\cite{CSGC2012} presents a bi-directional cooperative clustering model, where cluster members cooperate in inter-cluster and intra-cluster transmissions.  Both methods use a cost-sharing game in order to select CHs.  In~\cite{GTTOSN2013}, a coalitional game for heterogeneous WSNs is proposed.  In the proposed clustering method, a set of strong nodes, in terms of computing power and battery, are employed as the controllers of coalitions and their neighboring nodes form coalitions.  However, supporting heterogeneity and strong nodes is not practical in all setups. 

While prior algorithms utilize GT to alleviate the energy-efficiency problem of WSNs, they often fail to present a comprehensive design guideline on how to select essential network parameters.  Moreover, the majority of the algorithms rely on random cluster head election techniques, which, although is simple, considerably decreases the reliability, mainly because heads play the role of backbones in clustered networks and should have a high level of energy compared to other nodes.  Finally, none of the above works addresses the hotspot problem.  In contrast, we target the energy-efficiency, reliability and hotspot problems in WSNs through proposing CGTC, which is a deterministic clustering scheme utilizing coalitional games.

\section{Problem Formulation}
\label{net-mod}
We consider a network of $N$ nodes that are deployed in an area of size $M \times M$. The BS is located at a point far from the field.  The node positions follow a uniform random distribution so that we have
\begin{equation}
\label{eq:lambda}
N=\lambda |A|=\lambda M^2,
\end{equation}
where $\lambda$ is the node density and $A$ is the network area.  Both the BS and nodes are stationary, and the nodes are not equipped with GPS receivers.  All sensor nodes have the same capabilities and can use different power levels to communicate with other nodes.  The network operation is divided into rounds, where at the beginning of each round coalitions are formed and then data are disseminated to the BS through multi-hop paths among CHs. In-network data aggregation is applied to eliminate redundant sensor reports.  In the CGTC architecture, the network area is divided into two general regions: \textit{far} and \textit{vicinity}.  As mentioned earlier,  this division is adopted to solve the hotspot problem. We provide more details on this in the following sections.

The model for energy dissipation is derived from the radio model proposed in~\cite{Heinzelman2002}.  According to this model, the energy needed to transmit an $l$-bit packet to a node at distance $d$ is,
\begin{equation}
\label{eq:send-energy}
\pazocal{E}_{t}=
\begin{cases}
l(\pazocal{E}_{el} + \epsilon_{fs}d^{2}) \;\;\;\;\;\; d \leq d_{0} \cr
l(\pazocal{E}_{el} + \epsilon_{mp}d^{4}) \;\;\;\;\;\; d > d_{0}
\end{cases},
\end{equation}
where $\pazocal{E}_{el}$ is the electronics energy, $\epsilon_{fs}$ and $\epsilon_{mp}$ are the amplifier energy of free space and multi-path models, respectively, and $d_{0}=\sqrt{\epsilon_{fs}/\epsilon_{mp}}$. Also, to receive an $l$-bit packet, the energy consumed by a node is
\begin{equation}
\label{eq:receive-energy}
\pazocal{E}_{r}=l\pazocal{E}_{el}.
\end{equation}

 Considering $\Lb$ as the set of all players, any coalition $\pazocal{S}\subseteq \Lb$ stands for an agreement among players.  Moreover, $v$ indicates the worth of a coalition in a game.  Accordingly, a coalitional game is defined as follows.

\begin{defn}
A coalitional game is defined by the pair $(\Lb,v)$, where $\Lb$ is the set of players, and $v$ is the mapping function that determines the payoffs of players.
\end{defn}

The mapping function $v$, also called \textit{characteristic function}, is defined as $v: 2^\Lb \rightarrow \mathbb{R}$ and satisfies $v(\emptyset)=0$.  Coalitional games have two popular solutions: the \textit{core} and the \textit{Shapley value}.  In light of problems of the core solution~\cite{Han:2012:GTW}, we use the Shapley value.  We give the alternative interpretation of the Shapley value and the curious reader is referred to~\cite{shapley1952value,myerson1991game} for more information.  The payoff assigned to player \textit{i} is formulated as
\begin{equation}
\label{eq:shapley-value}
\phi_i(v)=\sum_{\pazocal{S}\subseteq \Lb \backslash \{i\}} \frac{|\pazocal{S}|!(\Lb-|\pazocal{S}|-1)!}{\Lb!}[v(\pazocal{S} \cup \{i\}) - v(\pazocal{S})],
\end{equation}
where $\phi$ is the payoff assigned in the game $(\Lb,v)$, and $|\pazocal{S}|$ is the cardinality of $\pazocal{S}$.  Since, in this paper, we focus on the \textit{cost games}, we use $c$ instead of $v$ in the following sections. 

The cost of each coalition is proportional to the amount of energy consumed by the coalition.  The consumed energy at each node has a direct relationship with receiving some data and transmitting it over a distance.  Depending on the location of nodes (i.e., located in the \textit{far} or \textit{vicinity} regions), the consumed energy is different.  Accordingly, the cost of a coalition located in the \textit{far} region is computed as
\begin{equation}
\label{eq:region-cost}
\pazocal{C}(\pazocal{S}^f)=\sum_{i\in \pazocal{S}^f} \pazocal{C}_{ch_i} + \sum_{j\in \pazocal{S}^f}\pazocal{C}_{nch_j},
\end{equation}
where $\pazocal{C}_{ch_i}(\pazocal{S}^f)$ is the consumed energy by a CH in coalition $\pazocal{S}^f$ and is found as
\begin{equation}
\label{eq:ch-cost}
\pazocal{C}_{ch_i}(\pazocal{S}^f)=lk(\pazocal{E}_{el} + \pazocal{E}_{da} + \mu(\pazocal{E}_{el}+\epsilon_{fs}d^{2})),
\end{equation}
where $\pazocal{E}_{da}$ indicates the energy dissipated for data aggregation, $k$ is the number of coalition members assigned to CH$_i$, and $\mu$ is the aggregation coefficient.  Note that here $d$ is the distance between a CH and its next-hop CH and could be $d^4$ if the distance is greater than $d_0$, based on Eq.~\eqref{eq:send-energy}.  Moreover, $\pazocal{C}_{nch_j}$ is the energy consumed by a non-CH node in coalition $\pazocal{S}^f$ and is calculated as 
\begin{equation}
\label{eq:member-cost}
\pazocal{C}_{nch_j}(\pazocal{S}^f)=l(\pazocal{E}_{el}+\epsilon_{fs}d^{2}).
\end{equation}
On the other hand, the cost of a coalition in the \textit{vicinity} region is proportional to the amount of energy that all ordinary nodes located within the same coalition spend to relay the received data.  Thus, 
\begin{equation}
\label{eq:vicinity-cost}
\pazocal{C}(\pazocal{S}^v)=\sum_{i\in \pazocal{S}^v} \pazocal{C}_{on_i},
\end{equation}
and
\begin{equation}
\label{eq:on-cost}
\pazocal{C}_{on_i}(\pazocal{S}^v)=l\pazocal{K}(\pazocal{E}_{el} + \pazocal{E}_{el} + \epsilon_{fs}d^{2})=l\pazocal{K}(2 \pazocal{E}_{el} + \epsilon_{fs}d^{2}),
\end{equation}
where $\pazocal{K}$ denotes the number of coalitions, located in the \textit{far} region, that select the coalition as their next-hop.

\section{Detailed CGTC Algorithm}
\label{sec:alg-det}

In this section, we explain  CGTC algorithm in detail.  The coalition formation in the \textit{far} region is performed with the objective of data gathering and transmitting to the BS.  On the other hand, in the \textit{vicinity} region, coalitions are formed in order to appropriately handle the data relying task imposed by farther coalitions.   Fig.~\ref{fig:tracking} illustrates a layout of CGTC architecture.  Note that the area division is performed once at the beginning of the network deployment by the BS through broadcasting a \textit{REG-DEC}, which is a light packet with few bytes about network partitioning, throughout the network. We first explain coalition formation in the \textit{far} region.

\subsection{Far Region Coalition Formation}
\label{subsec:far-coal}
Our main objective of coalition formation is energy efficiency so the remaining energy plays a crucial role in selecting head nodes.  
At the beginning of each round, all the nodes broadcast a \textit{CHN-INF} packet containing information about the residual energy  ($E_{res}$), the number of neighbors (node degree, $N_d$), and the proximity to the BS ($d_{i,bs}$), within their \textit{coalition range ($R_c$)}. Then, they wait for a predefined time ($T_{c}$), which is a function of $R_c$ and node density.  Having waited for $T_{c}$ and received the packets from neighbors, each node either selects itself as the CHN based on the highest residual energy or remains the ordinary node. Here the ties are randomly broken in the first round. 

\begin{figure}[t]
\centering
	\subfigure[Network architecture]{
	\includegraphics[width=0.4\linewidth]{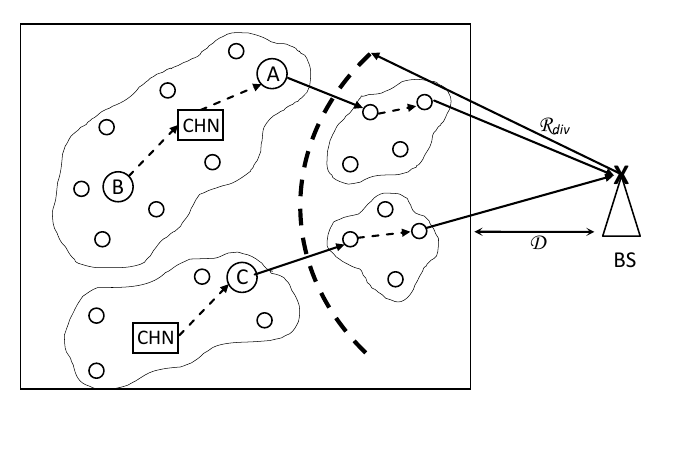}
    \label{fig:tracking}
	}
	\subfigure[Routing]{
	\includegraphics[width=0.4\linewidth]{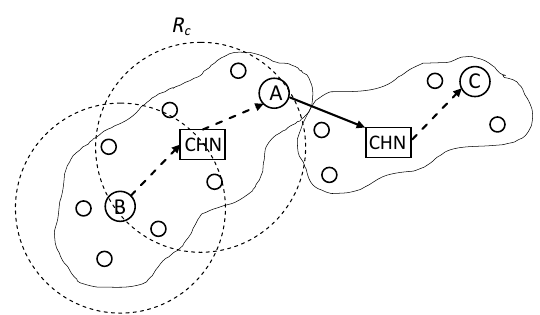}
    \label{fig:routing}
	}
\caption{The network architecture and routing scheme in CGTC}
\label{fig:Rdiv}
\end{figure}

When CHNs are elected, they select one or two other nodes in order to constitute their coalitions.  CHNs select two nodes with the highest $N_d$ and smallest $d_{i,bs}$, that are located within the $R_c$ of CHNs, as their Coalition Node degree Ally (CNA) and Coalition Distance Ally (CDA), respectively.   To do so, the CHN unicasts a \textit{CH-REC} packet to potential CHs.  The condition for cooperation is that the initial cost of the CHN should be reduced.  In other words, if there is a node, e.g. A, within the $R_c$ of the CHN, the CHN cooperates with A if $\phi_{CHN} + \phi_A < \pazocal{C}(\{CHN\})$, where  $\pazocal{C}(\{CHN\})$ is the cost of coalition when $\pazocal{S}=\{CHN\}$ and is computed using Eq.~\eqref{eq:region-cost}.  Note that each node computes its share using Eq.~\eqref{eq:shapley-value}. 
   Therefore, the final successful coalition in this example has three CHs, i.e., $\pazocal{S}=\{CHN, A, B\}$ (see Fig.~\ref{fig:tracking}).

Elected as CHs, head nodes broadcast an advertisement message \textit{CH-ADV} within their $R_t$ and invite ordinary nodes to become their coalition members.  The reason why CHs broadcast \textit{CH-ADV} within their $R_t$ instead of $R_c$ is that they can use this message to find next-hops in the routing process, which is described in the balance of this subsection.
On receiving this message, each non-head node sends a \textit{JOIN-REQ} message to the nearest head node based on Received Signal Strength Indicator (RSSI)~\cite{Heinzelman2002}.  Then, CHs establish a Time Division Multiple Access (TDMA) protocol and send time schedules to their members. 

Once coalitions are formed, data should be transmitted to the BS periodically.  To do this, multi-hop paths are established among CHs.    The routing task in our scheme is  divided into: intra-coalition and inter-coalition routing.  Intra-coalition routing is performed during coalition formation.  The next-hop of a CHN is always a CDA while it is possible for a CNA not to be within the same coalition range with CDA (e.g. nodes A and B in Fig.~\ref{fig:routing}).  In such a case, the CNA forwards its data to the CHN, which relays the data to the CDA.  In inter-coalition routing, each CDA transmits its data to the nearest head nodes of other coalitions that are closer to the BS, irrespective of their type (CHN, CNA, or CDA).  As mentioned earlier, CDAs find the closest CHs from other coalitions by hearing their \textit{CH-ADV} and calculating the distance by RSSI.  Therefore, the data is relayed among head nodes toward the BS to reach the \textit{vicinity} area, where the main factor for routing is the distance and nodes find their next-hop based on shortest-path.
 
\subsection{Vicinity Region Coalition Formation}
\label{subsec:vic-reg}
In the \textit{vicinity} region, there is no CH and these are boundary nodes that usually start the game.  When boundary nodes receive data from far coalitions, they calculate the cost of direct transmission to the BS.  Then, they select up to three nodes, preferably with high $E_{res}$, and check the condition of cooperation, explained in the previous subsection.  The reason why boundary nodes select up to three nodes for coalition formation is to keep coalitions small enough and the fact that computing the Shapley value for a larger number of players becomes more complex~\cite{GT-survey2015}.
Note that since the defined cost is in fact the energy  consumed by a coalition, if nodes that join the coalition have a higher $E_{res}$ than the node that starts the game, this extra energy is subtracted from the coalition cost, which is computed by Eq.~\eqref{eq:vicinity-cost}.
Note also that nodes' priority in the {\it vicinity} region is the data relaying task.
The pseudo code of coalition formation in CGTC is presented in Algorithm~\ref{alg:ch-el}.

\begin{algorithm}
\label{alg:ch-el}
\caption{ Coalition formation in CGTC}
\SetAlgoLined
\DontPrintSemicolon

\While {$ i < N $}
{
broadcast {\it CHN-INF} within $R_{c}$\;
wait for $T_{c}$ to receive {\it CHN-INF}\;
\eIf{ $\forall j, E_{res}(i)=E_{res}(j)$ }{
check the coalition condition and unicast {\it CH-REC} to CNA and CDA\;}
{wait $T_{c}$ for {\it CH-ADV}}

\eIf{ $i$ = {\upshape CH} }{
broadcast {\it CH-ADV} within $R_{t}$\;}
{send \textit{JOIN-REQ} to the nearest CH\;}
}
\end{algorithm}

\section{Analysis}
\label{sec:anal}

\begin{figure}[t]
\centering
	\includegraphics[width=.4\linewidth]{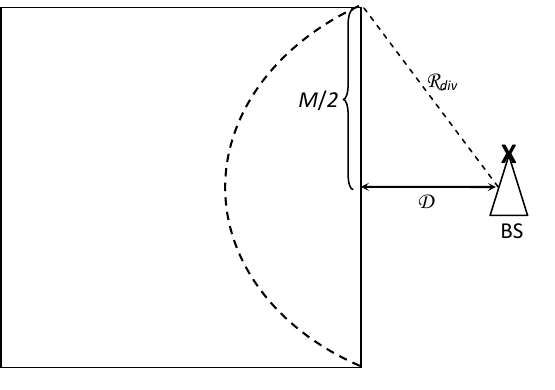}
\caption{The division range ($\pazocal{R}_{div}$) analysis }
\label{fig:division}
\end{figure}

In this section, due to the limited space, we suffice to analyze $\pazocal{R}_{div}$, the radius that divides the \textit{far} and \textit{vicinity} regions.  In general, $\pazocal{R}_{div}$ should not be too small, as the network might become disjointed if some set of nodes die, and should not be too large either, otherwise the network becomes a flat architecture.  One way to select the value of $\pazocal{R}_{div}$ is to use heuristic algorithms, which are not practical in all WSNs' setups mainly because they are very time and energy consuming. In this paper, we give an approximate but realistic $\pazocal{R}_{div}$ and leave the optimal $\pazocal{R}_{div}$ for a future contribution.  We first define the upper and lower bounds and give an approximate $\pazocal{R}_{div}$. Then, using simulation, we examine our approximation. The upper bound equals the transmission range ($R_t$) of a node, because if there is only one node in the \textit{vicinity} region, it should be able to directly transmit the received data to the BS.  On the other hand, the lower bound is the distance between the network and the BS ($\pazocal{D}$) plus a range wherein there is at least one sensor node.  We call this range $R_{min}$ and, based on Eq.~\eqref{eq:lambda}, is calculated as
\begin{equation}
\label{eq:r-min}
R_{min} =\sqrt{\frac{1}{\lambda \pi} }.
\end{equation}
From Eq.~\eqref{eq:r-min}, it is observed that $R_{min}$ depends on the node density.  Thus, considering $\pazocal{D} < R_t$ (since the network becomes disjointed otherwise),
\begin{equation}
\label{eq:radius_region}
\pazocal{D} + R_{min} \leq \pazocal{R}_{div} \leq R_t.
\end{equation}
$R_t$ is a very important factor to keep the network connected. As discussed and proven in~\cite{Younis2004,ACCA}, in order to form a connected network topology, $R_t\geq 6R_c$.  In our simulations, $R_t=6R_c$.   A simple yet effective approximation for $\pazocal{R}_{div}$ could be the distance between the top right playfield's corner and the BS (see Fig.~\ref{fig:division}).  In other words, $\pazocal{R}_{div}$ may be calculated as
\begin{equation}
\label{eq:radius_region1}
\pazocal{R}_{div} = \sqrt{\pazocal{D}^2 + (M/2)^2}.
\end{equation}
We will discuss this in the next section, where we provide experiments for different values of $\pazocal{R}_{div}$ and its impact on the performance of the network.

\section{Performance Evaluation}
\label{sec:per-eval}
In this section, the results of our experiments through simulations are given.
\subsection{Simulation Setup}
\label{subsec:sim-set}
Four different node densities are studied: $\lambda_1=0.02$ ($N=200, M=100$),  $\lambda_2=0.005$ ($N=M=200$), $\lambda_3=0.0022$ ($N=200, M=300$), and $\lambda_4=0.00125$ ($N=200, M=400$), based on Eq~\eqref{eq:lambda}.  
The BS is located at $(M+50,M/2)$ and the nodes are dispersed using a uniform random distribution.  We take the duration of each round to be equal to five data gathering epoch.  For simplicity, we do not consider packet losses and assume that all messages are successfully received by their destinations. 
 In order to measure the overhead of cluster formation and routing, we assume that each packet has a two-byte header and one byte trailer~\cite{2003energy}.  The payload length of a packet depends on the message type.

We compare CGTC with two baseline clustering algorithms, namely MLEACH~\cite{Mhop-leach} (the multi-hop version of LEACH protocol~\cite{Heinzelman2002}) and CGC~\cite{intech2010}.  In general, we have picked the parameter settings that yield the best performance for baseline algorithms.  In particular, for both competing algorithms, depending on the node density, we have the CH election probability $p\in[0.05,0.1]$.  Other simulation parameters are summarized in Table~\ref{tab:simpar}.  The individual results are the average over 50 runs and the length of each run is $r=10000$ rounds.  When subjected to 95\% confidence interval the results stayed within 6-10\% of the sample mean. 
 
\begin{table}

\caption{Simulation Parameters}
\centering
\tiny
\begin{tabular}{| l || c |}
\hline
{\bf Parameter} & {\bf Value} \\ [0.5ex]
\hline
\hline
$N$ & 200 \\
\hline
$M$ & $100\sim400$ \\
\hline
BS & (M+50, M/2)\\
\hline
$\epsilon_{fs}$ & $10pJ/bit/m^2$\\
\hline
$\epsilon_{mp}$ & $0.0013pJ/bit/m^4$\\
\hline
$\pazocal{E}_{el}$ & $50nJ/bit$ \\
\hline
$\pazocal{E}_{da}$ & $5nJ/bit/signal$\\
\hline
$\mu$ & 0.5\\
\hline
Initial Energy & $5J$ \\
\hline
Data Payload & $100$B\\
\hline
Header & $2$B\\
\hline
Trailer & $1$B\\
\hline
\end{tabular}
\label{tab:simpar}
\end{table}

\subsection{Simulation Results}
\label{subsec:sim-res}
\begin{figure}[]
\setlength\abovecaptionskip{-0\baselineskip}
\centering
	\subfigure[ $\lambda_1$]{
	\includegraphics[width=0.4\linewidth]{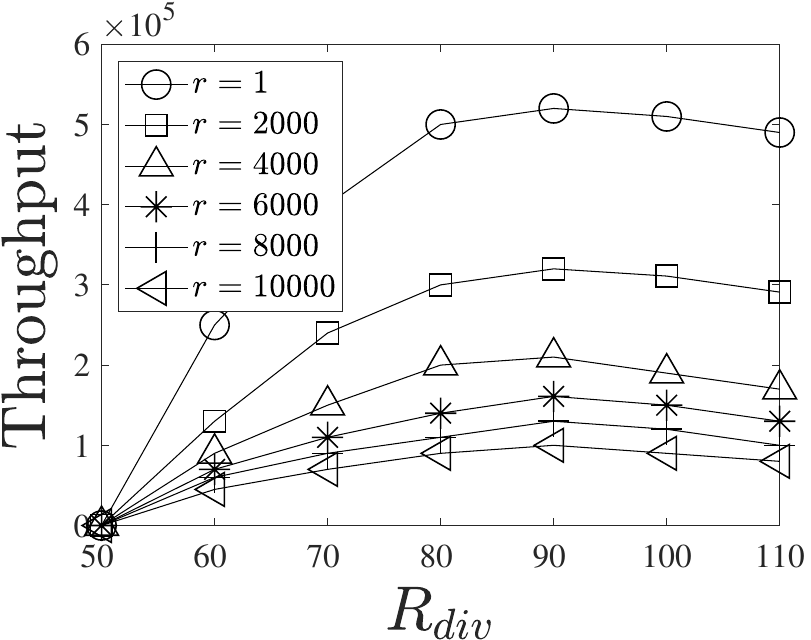}
    \label{fig:rdive1}
	}
	\subfigure[ $\lambda_2$]{
	\includegraphics[width=0.4\linewidth]{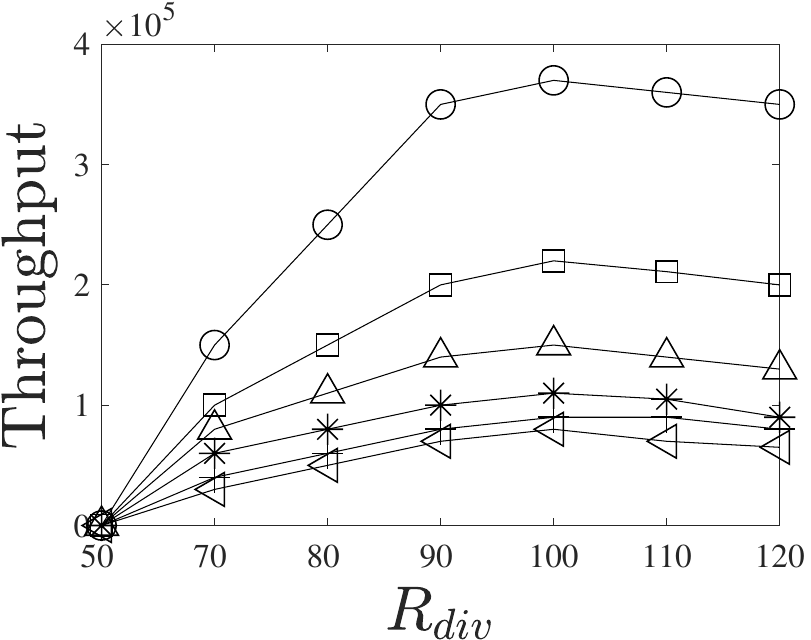}
    \label{fig:rdive2}
	}
	\subfigure[ $\lambda_3$]{
	\includegraphics[width=0.4\linewidth]{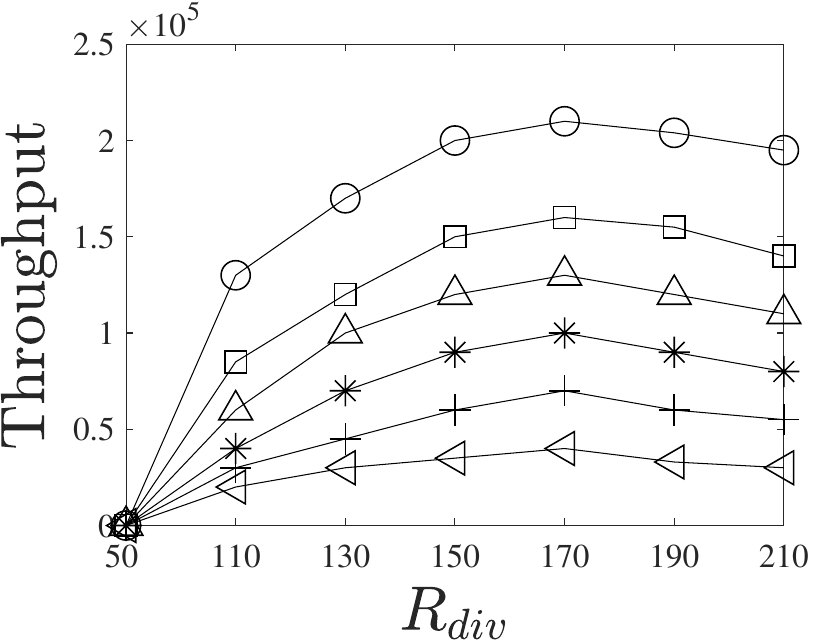}
    \label{fig:rdive3}
	}
	\subfigure[ $\lambda_4$]{
	\includegraphics[width=0.4\linewidth]{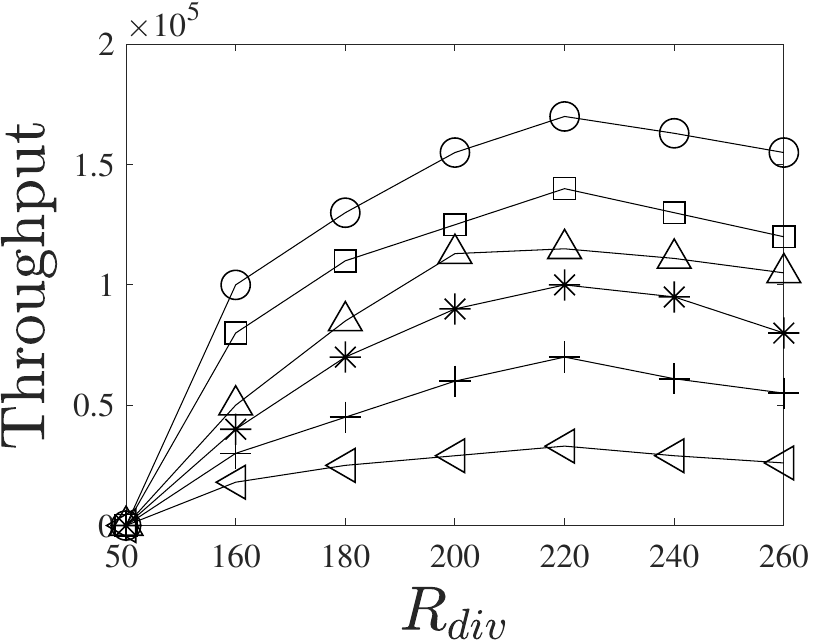}
    \label{fig:rdive4}
	}
\caption{The system throughput in CGTC when $\pazocal{R}_{div}$ and $\lambda$ vary}
\label{fig:Rdiv}
\end{figure}

In this section, the results of simulations are analyzed and interpreted. We first evaluate the impact of $\pazocal{R}_{div}$ on the throughput, which is the number of data bits reaching the BS successfully. According to Fig.~\ref{fig:Rdiv}, for all densities studied, the throughput increases as $\pazocal{R}_{div}$ grows and reaches a peak somewhere in the middle, and then becomes stable or diminishes slightly.  For example, in $\lambda_1$, while the throughput reaches it maximum at $\pazocal{R}_{div}=80m$, it slightly degrades when $\pazocal{R}_{div}>90m$.  
This number for  $\lambda_2$, $\lambda_3$, and $\lambda_4$ are 100, 170, and 220$m$, respectively, depicted in Figs.~\ref{fig:rdive2},~\ref{fig:rdive3}, and~\ref{fig:rdive4}.  This is consistent with our analyses performed in section~\ref{sec:anal}.  According to Eq.~\eqref{eq:radius_region1}, $\pazocal{R}_{div}$ for four node densities are 70, 111, 158, and 206$m$.  Thus, we can conclude that Eq.~\eqref{eq:radius_region1} provides a good approximation for the value of $\pazocal{R}_{div}$.

\begin{figure}[]
\centering
	\subfigure[ $\lambda_1$]{
	\includegraphics[width=0.4\linewidth]{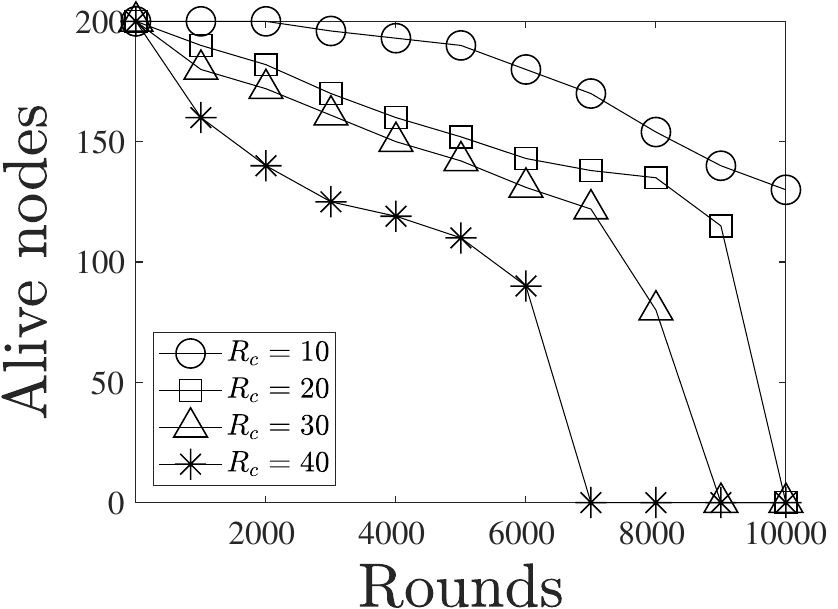}
    \label{fig:life1}
	}
	\subfigure[$\lambda_2$]{
	\includegraphics[width=0.4\linewidth]{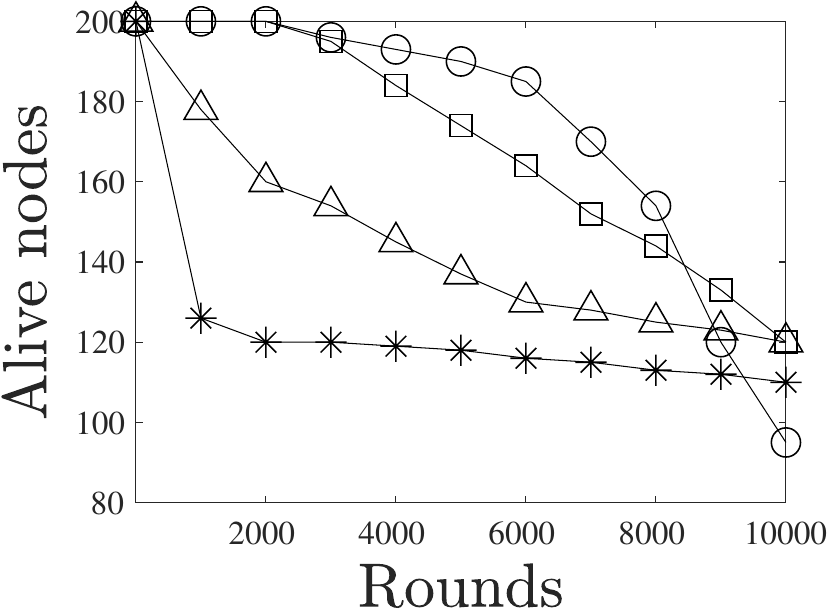}
    \label{fig:life2}
	}
	\subfigure[$\lambda_3$]{
	\includegraphics[width=0.4\linewidth]{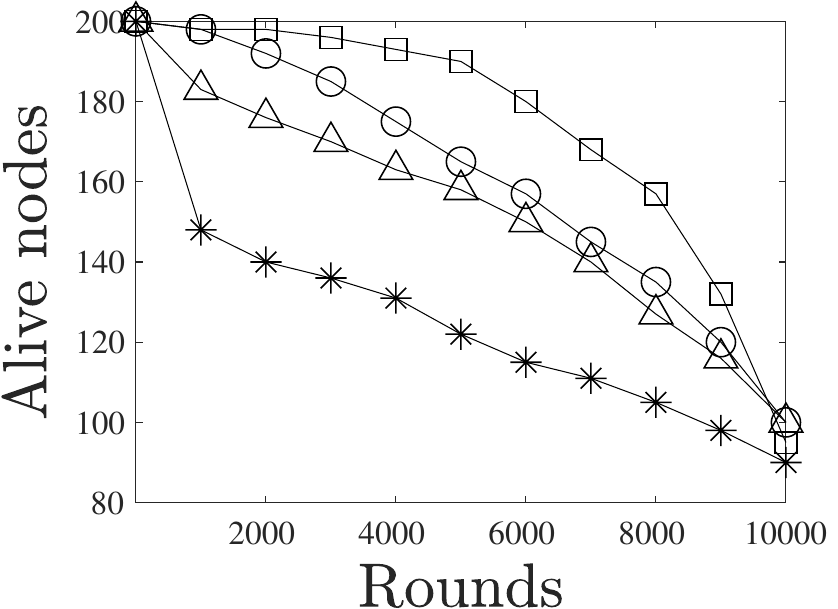}
    \label{fig:life3}
	}
	\subfigure[$\lambda_4$]{
	\includegraphics[width=0.4\linewidth]{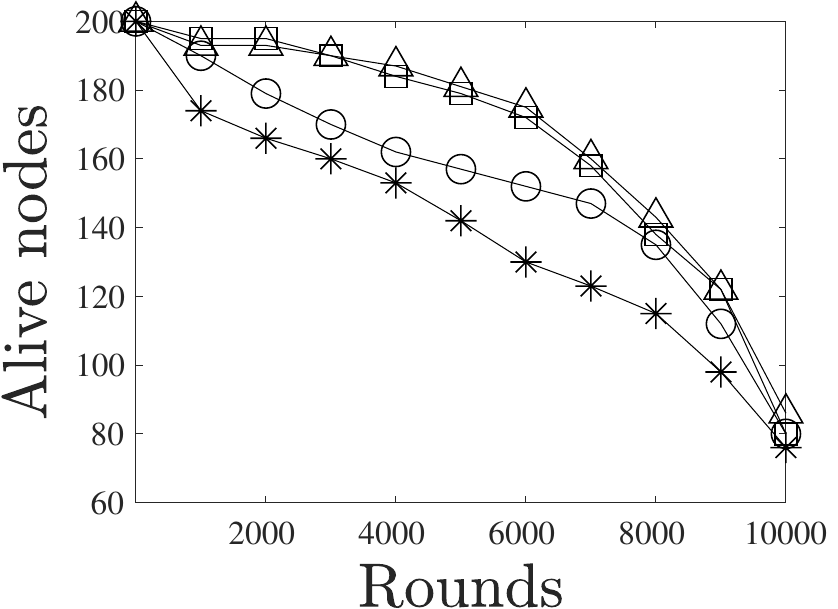}
    \label{fig:life4}
	}
\caption{The number of alive nodes in CGTC when $R_c$ and $\lambda$ vary}
\label{fig:life-own}
\end{figure}

 Fig.~\ref{fig:life-own} illustrates the number of alive nodes when $R_c$ varies.  As shown in Fig.~\ref{fig:life1} for $\lambda_1$, CGTC has the best performance when $R_c=10m$.  This is mainly because when the network is small ($M=100$), smaller coalitions are more suitable and the network's energy consumption diminishes.  On the other hand, when the network grows bigger, a larger $R_c$ works better. For example, in $\lambda_3$, the number of alive nodes is the best when $R_c=20m$.  This is consistent with our analysis, discussed in section~\ref{sec:alg-det}, that $R_c$ is a function of the node density.

\begin{figure}[]
\centering
	\subfigure[ $\lambda_1$, $R_c=10m$]{
	\includegraphics[width=0.4\linewidth]{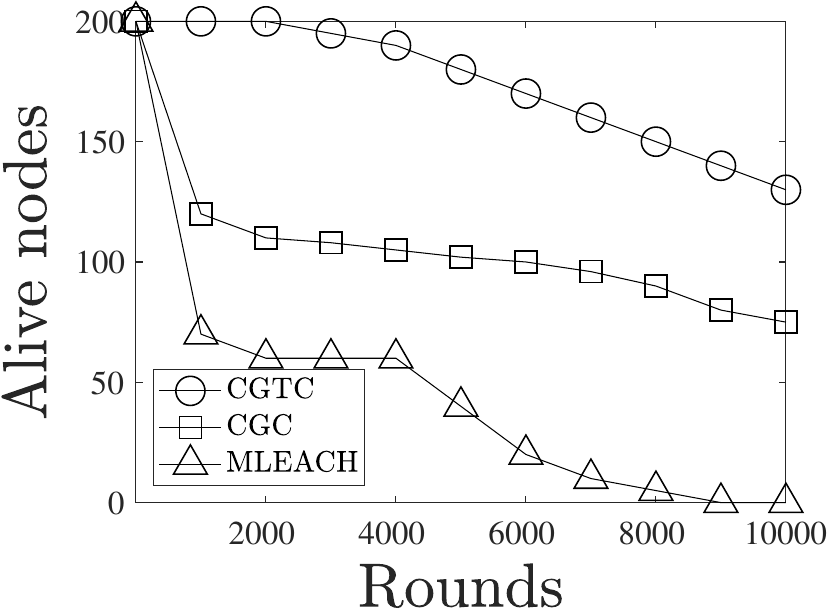}
    \label{fig:100node}
	}
	\subfigure[$\lambda_2$, $R_c=20m$]{
	\includegraphics[width=0.4\linewidth]{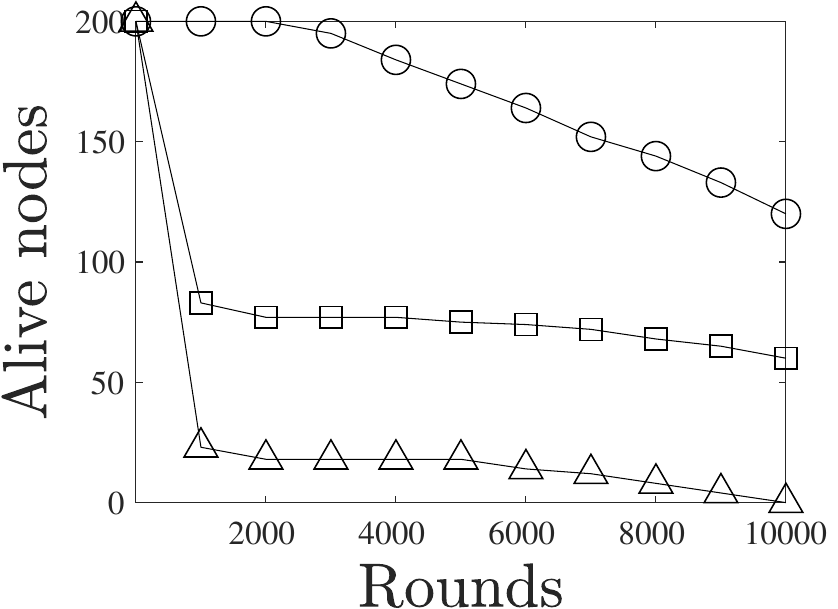}
    \label{fig:200node}
	}
	\subfigure[$\lambda_3$, $R_c=20m$]{
	\includegraphics[width=0.4\linewidth]{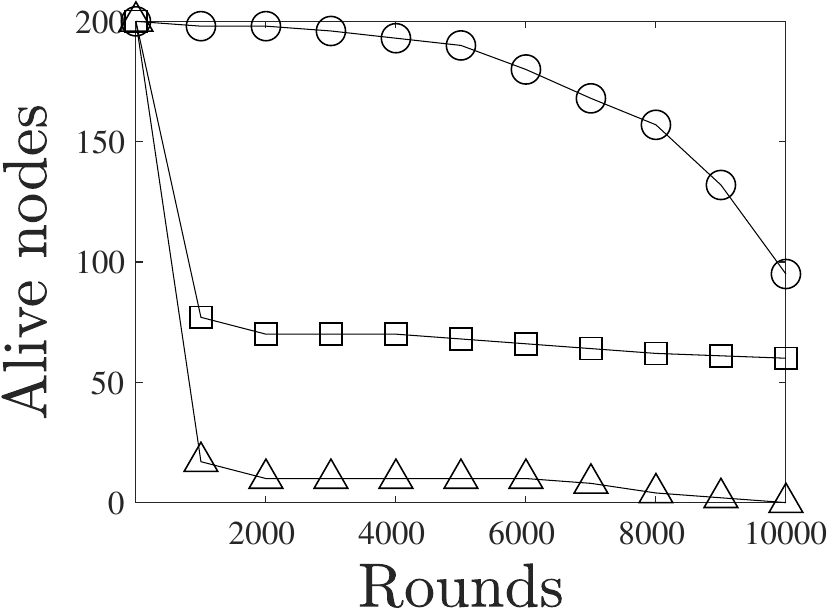}
    \label{fig:300node}
	}
	\subfigure[$\lambda_4$, $R_c=20m$]{
	\includegraphics[width=0.4\linewidth]{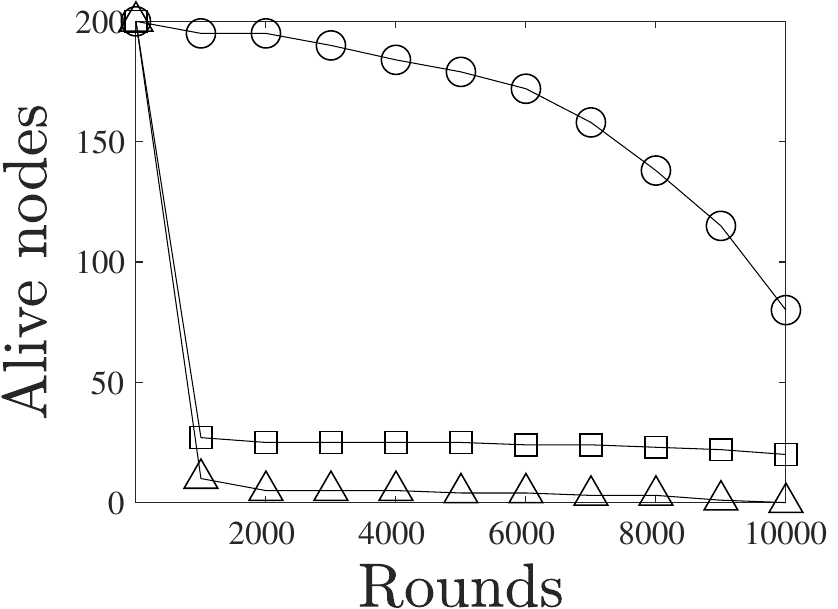}
    \label{fig:400node}
	}
\caption{The number of alive nodes in the three algorithms when $\lambda$ changes}
\label{fig:life}
\end{figure}

 Fig.~\ref{fig:life} compares the number of alive nodes for the three algorithms when $\lambda$ varies.  As shown in Fig.~\ref{fig:100node}, CGTC significantly outperforms the two competing algorithms in terms of the number of alive nodes.  While in CGTC all nodes are alive until round 2000, it experiences a relatively constant decrease, such that more than 60\% of nodes are alive at the end (round 10000).  However, in CGC, the number of alive nodes quickly drops from 200 to around 100 in the first 1000 rounds, and it then slightly decreases until it reaches 70 at the end.  Likewise, in MLEACH, nodes die more rapidly and around two-third of them die within the first 1000 rounds, and after being stable at 50 till round 4000, the number of alive nodes considerably drops and all nodes die before round 9000.  On the other hand, according to Figs.~\ref{fig:200node} to~\ref{fig:400node}, while CGTC suitably handles the scalability problem, the performance of the competing algorithms significantly degrades when the network becomes larger.  For instance, as shown in Fig.~\ref{fig:400node}, although more than 90\% of all nodes are alive in CGTC until round 5000, the number of alive nodes in CGC and MLEACH is around 25 and 5, respectively. The reason behind this superiority is that CGTC forms deterministic coalitions, picks nodes with appropriate parameter values as heads, and solves the hotspot problem.   Among baseline algorithms, CGC has a better performance, as it employs coalitional games; however, the algorithm still suffers from probabilistic head election, inefficient routing, and hotspot problem. 

\begin{figure}[]
\centering
	\subfigure[ $\lambda_1$, $R_c=10m$]{
	\includegraphics[width=0.4\linewidth]{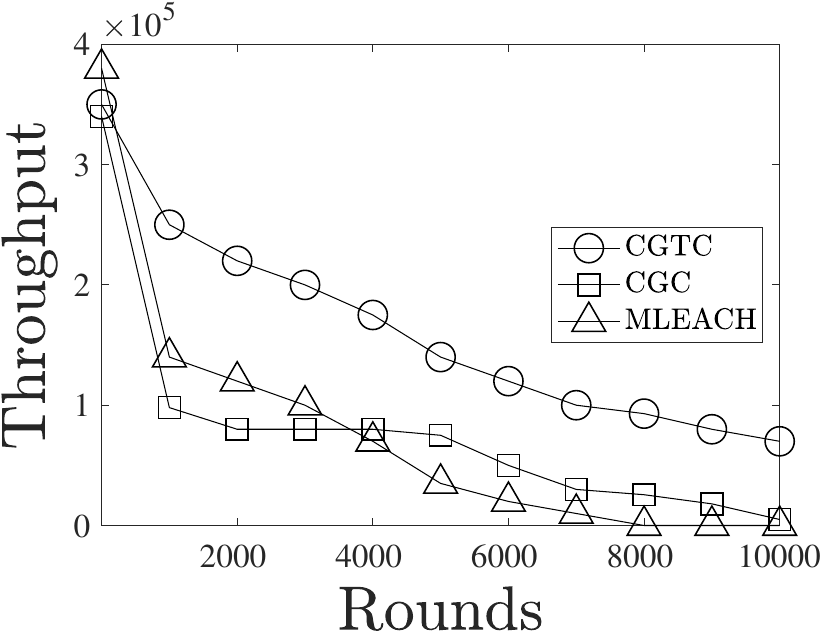}
    \label{fig:thr100}
	}
	\subfigure[ $\lambda_2$, $R_c=20m$]{
	\includegraphics[width=0.4\linewidth]{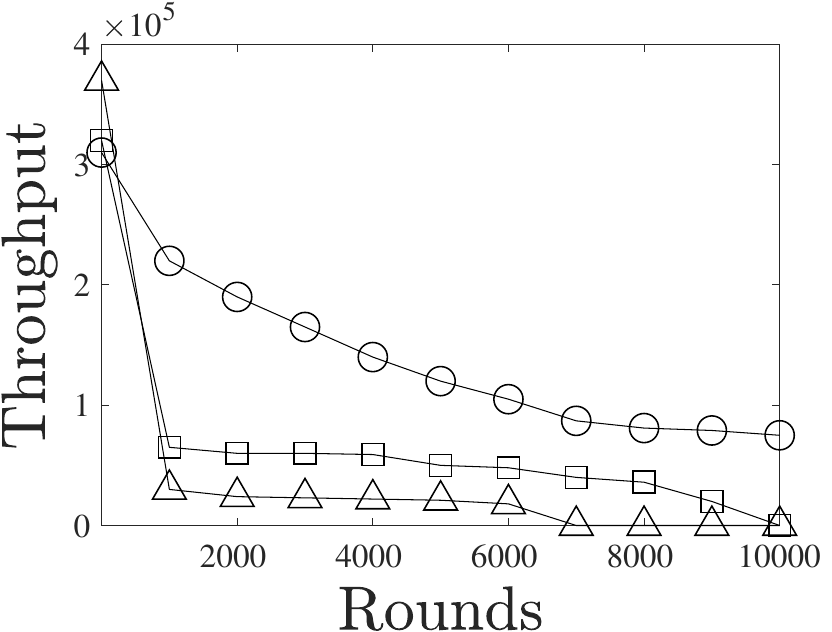}
    \label{fig:thr200}
	}
	\subfigure[ $\lambda_3$, $R_c=20m$]{
	\includegraphics[width=0.4\linewidth]{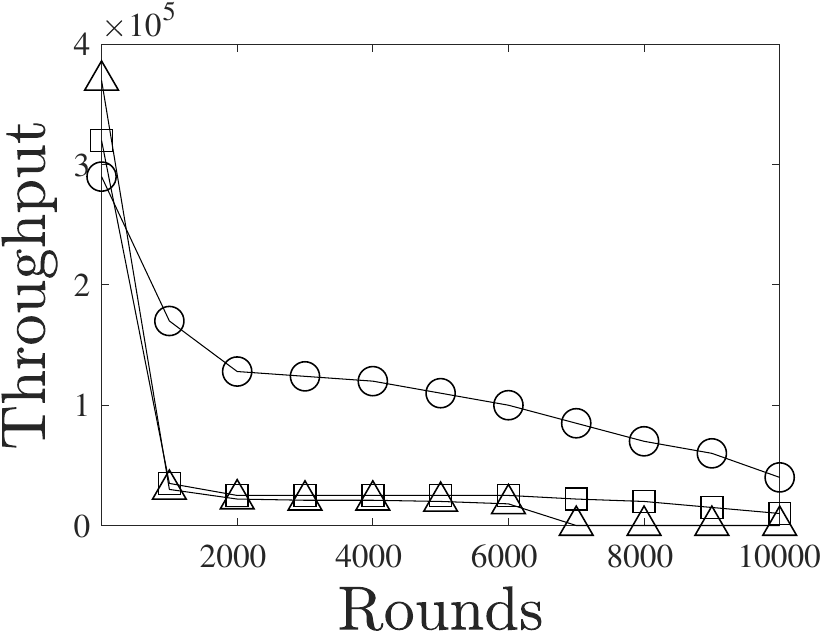}
    \label{fig:thr300}
	}
	\subfigure[ $\lambda_4$, $R_c=20m$]{
	\includegraphics[width=0.4\linewidth]{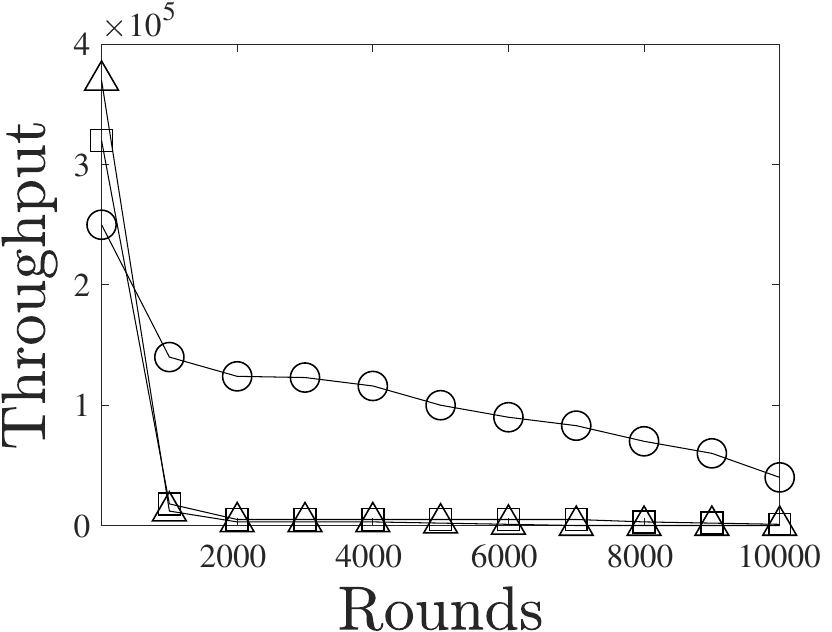}
    \label{fig:thr400}
	}
\caption{The system throughput in the three algorithms when $\lambda$ changes}
\label{fig:thr-cmp}
\end{figure}

Finally, the throughput of the system is shown in Fig.~\ref{fig:thr-cmp} for the three algorithms.  The data indicates that the CGTC's throughput is far better than that of the two baseline algorithms.  For example in $\lambda_4$, when the effective throughput in MLEACH and CGC is almost zero, it is higher than $1\times10^5$ in CGTC before round 5000.  
This superiority is mainly because CGTC has a good network lifetime and preserves a suitable level of network connectivity.

 It is noteworthy to mention that we have achieved the results above for a homogeneous WSN, but this might not be the case for a heterogeneous WSN.  Moreover, we have not considered packet loss in the implementation. While it might impact the network behavior and achieved results, it is not an unrealistic assumption to make. This is because the main source of packet loss is collision and since CGTC utilizes TDMA protocol and message redundancy, intra- and inter-coalition node interference is minimized~\cite{heed}.

\section{Conclusion}
\label{sec:con}
We have proposed CGTC, a game-theoretic clustering algorithm for sensor networks, whose main objective is energy-efficiency. CGTC is a distributed, reliable clustering scheme which solves the hotspot problem in WSNs through forming small coalitions close to the BS and larger coalitions as the distance to the BS increases. The results of simulations show that CGTC outperforms competing algorithms in terms of energy-efficiency and system throughput. In future, we plan to mathematically optimize the division range and to study its effect on the performance of CGTC.

\section*{Acknowledgment}
This paper was partially supported by NSERC (Natural Sciences and Engineering Research Council of Canada).

\bibliographystyle{IEEEtran}
\bibliography{general}

\end{document}